\documentclass[prl,aps,twocolumn]{revtex4}
\usepackage{graphicx,color,latexsym}
\usepackage{dcolumn}
\usepackage{amsmath,amssymb,epsf,bm}
\begin{document}
\title{Magnon-induced transparency of a disordered antiferromagnetic Josephson junction}
\author{A.~G. Mal'shukov}
\affiliation{Institute of Spectroscopy, Russian Academy of Sciences, Troitsk, Moscow, 108840, Russia}

\begin{abstract}
We considered a planar Josephson junction which is composed of two s-wave superconducting contacts deposited on the top of a  thin antiferromagnetic (AFM) disordered metal film. In such a system noticeable Josephson currents  may be observed, if contacts are just nanometers away from each other. It is shown that the excitation of AFM by magnons results in  a strong enhancement of the stationary current through much longer junctions, whose length may be comparable to the coherence length of superconducting correlations in a nonmagnetic metal. Such a current is calculated at the weak tunneling amplitude of electrons between superconducting contacts and AFM. The problem is considered for nonequilibrium Green functions in the second-order perturbation theory with respect to the electron-magnon interaction. A spin-orbit torque oscillator was taken as a possible source of long-wavelength classic magnetic waves. This work predicts a strong effect of magnons on superconducting proximity effect in AFM, with promising applications in superconducting spintronics.
\end{abstract}
\maketitle

\emph{Introduction}-The interplay of superconductivity and magnetism results in a number of physical phenomena which are important for fundamental research and applications in many fields of solid state physics \cite{Bergeret,Buzdin,Robinson,Qi,Fu}. Most of research was focused on studying systems which combine ferromagnetic (FM) and superconducting (S) materials in nanosized heterostructures. On the other hand, recently, AFM materials attracted a special attention, as a possible component of such structures. AFM may efficiently transfer the orbital moment through interfaces of AFM with other magnetic and nonmagnetic materials \cite{Baltz,Gomonay,Yan,Wadley}. In comparison to ferromagnets they can operate much faster in transmitting spin currents by magnons over large distances \cite{Cornelissen,Lebrun}. Besides, AFM's do not produce stray magnetic fields which can induce undesirable effects in heterostructures containing superconductors. A key mechanism, which operates near AFM-S, or FM-S interfaces, is the superconducting proximity effect. It is caused by penetration of electron's pair correlations from S into a contacting magnetic material. The penetration depth from an s-wave spin-singlet superconductor into a strong FM is very small (less than nm)  \cite{Bergeret,Buzdin}. Surprisingly, it is also small in a disordered AFM \cite{Krivoruchko,Bobkov,Brataas,Bell,Weides}, even though the macroscopic magnetic field is absent there. On that reason, the Josephson effect, where a disordered AFM is used as a weak link, is strongly suppressed when the distance between superconducting contacts exceeds the electron's mean free path  and the exchange interaction of conduction electrons with localized spins is sufficiently strong.

The suppression of  the proximity effect occurs when an AFM makes a contact with a spin-singlet superconductor. On the other hand, as noted in \cite{Brataas}, the long-range proximity effect
takes place for triplet Cooper pairs whose spins are perpendicular to the N\'{e}el order. Such a penomenon is well known for ferromagnets with an inhomogeneous magnetization \cite{Bergeret2,Kadigrobov}. In Josephson junctions (JJ) such triplets can be created by ferromagnetic interlayers placed between superconductors and AFM \cite{Malsh}. Other ways for creating the triplet pairing in proximized FM and AFM were considered in \cite{Linder,Rezaei,Chaou}. A  new opportunity for creating and detecting  of triplet superconducting correlations in AFM opens, if one takes into account that triplets carry the angular moment which, in turn, can interact with magnetic excitations of the N\'{e}el order.  In which connection, we considered the effect of magnons on the critical current through a phase-biased S-AFM-S planar junction. The idea is that magnons can transfer their angular moment to electrons. By flipping electron spins magnons promote a transformation of singlet pairing correlations into triplet ones. Due to this conversion magnetic excitations lead to a strong increase of the critical current through a disordered JJ whose length is larger than the electron's mean free path, because triplet correlations can extend over  large distances. Such a strong effect of magnons on the Josephson current opens the way for new spintronic applications in hybrid systems which combine superconducting circuits and magnetic materials.

\begin{figure}[bp]
\includegraphics[width=6cm]{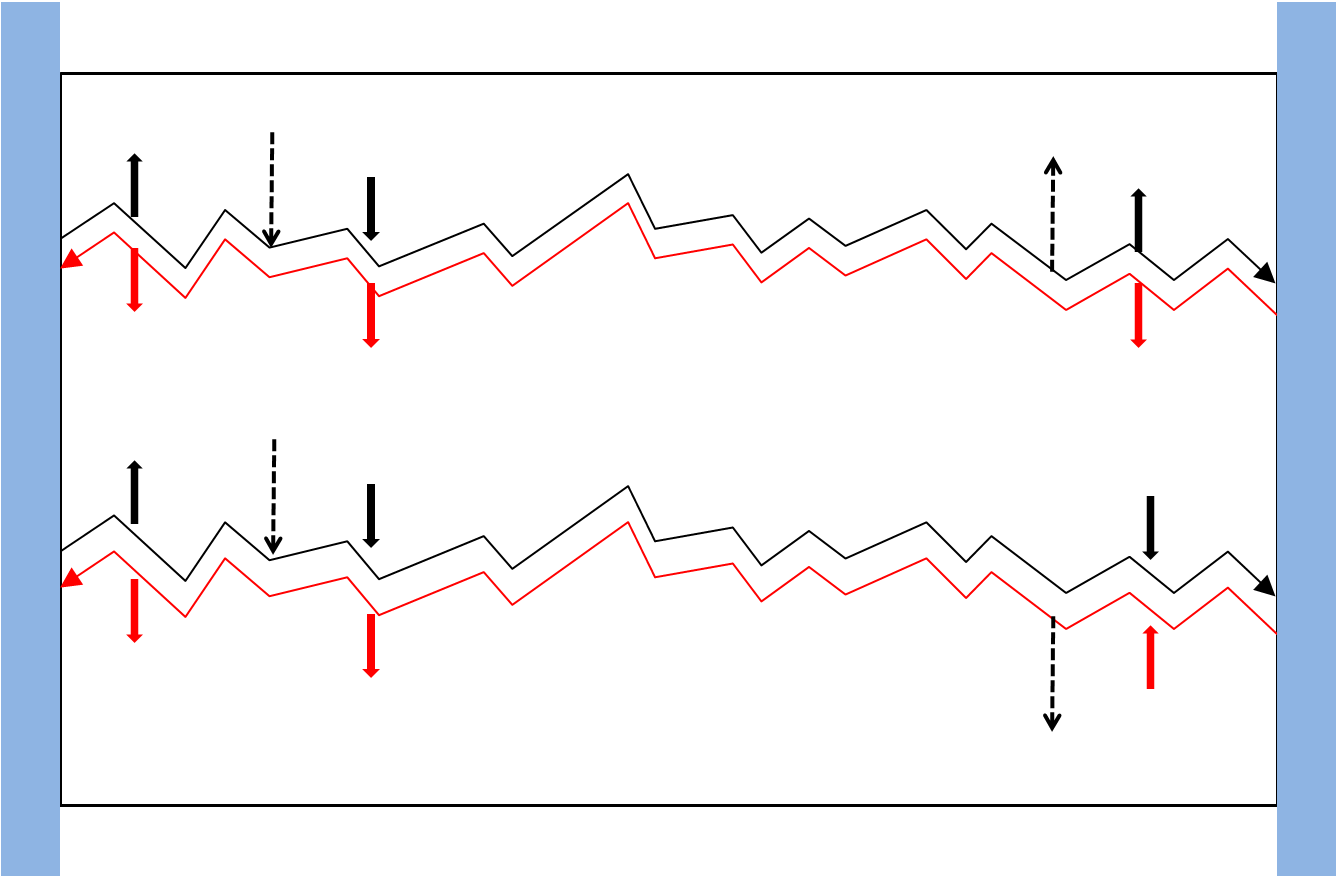}
\caption{(Color online) Two possible random walk paths of correlated electrons and Andreev-reflected holes between superconducting contacts. Dashed arrows depict absorbed and emitted magnons, while thick arrows show electron and hole spins. The interaction with magnons results in flipping spin's directions. Therefore, the total two-particle spin switches between singlet and triplet states. The latter may diffuse over relatively large distances} \label{fig1}
\end{figure}
The main physics of the considered effect is explained in Fig.1. The dc current is driven by the phase bias. Two-magnon absorption-emission processes are taken into account perturbatively. For simplicity, the coherent excitation of the N\'{e}el order  is treated as a classical magnetic wave whose length is much larger than the distance between S-contacts. Magnons may be excited and transmitted to JJ by many ways. Here, it is assumed that the magnetic excitation is produced by the spin-orbit torque oscillator \cite{Cheng,Khymyn,Khymyn2,Liu}. The corresponding setup is discussed below in Supplemental material (SM).

\emph{Formalism}-We assume a weak coupling between an AFM and a superconductor, where the latter is characterized  by the s-wave singlet order parameter. The AFM is represented by a thin film with superconducting contacts deposited on its top. The corresponding tunneling Hamiltonians $H_{L}$ and  $H_{R}$  for the left and right contacts can be written in the form
\begin{equation}\label{Hint}
H_{L(R)}=\sum_{\mathbf{k},\mathbf{k}^{\prime}\lambda}(\psi^{\dag}_{\lambda\mathbf{k}}t^{\lambda}_{L(R)\mathbf{k},\mathbf{k}^{\prime}}
\tau_3\psi^{s}_{L(R)\mathbf{k}^{\prime}}+ h.c.)\,,
\end{equation}
where the field operators of AFM electrons  are defined in the Nambu basis in terms of  creation and destruction operators with the wave vector $\mathbf{k}$ and spin projections $\uparrow$ and $\downarrow$, as $\psi_{\lambda\mathbf{k}}=(c_{\lambda\mathbf{k}\uparrow},c_{\lambda\mathbf{k}\downarrow},c^{\dag}_{\lambda\mathbf{-k}\downarrow},
-c^{\dag}_{\lambda\mathbf{-k}\uparrow})$. The field operators $\psi^{s}_{L(R)\mathbf{k}^{\prime}}$ in superconducting contacts are defined in a similar way. The Pauli operators $\tau_i$, with $i\in(1,2,3)$, operate in the Nambu space and $\lambda \in (1,2)$ is the sublattice variable, with the corresponding Pauli operators $\lambda_l$, $l\in (x,y,z)$. Tunneling amplitudes $t^{\lambda}_{L(R)\mathbf{k},\mathbf{k}^{\prime}}$ are assumed to be spin-independent. By commuting Eq.(\ref{Hint}) with the total number of electrons in the superconductor it is easy to find the current operators for the left and right interfaces with AFM \cite{Aslamazov}:
\begin{equation}\label{jhat}
\hat{j}_{L(R)}=ie\sum_{\mathbf{k},\mathbf{k}^{\prime}\lambda}(\psi^{\dag}_{\lambda\mathbf{k}}t^{\lambda}_{L(R)\mathbf{k},\mathbf{k}^{\prime}}\psi^{s}_{L(R)
\mathbf{k}^{\prime}}- h.c.)\,.
\end{equation}
The observable current can be calculated by using the disorder averaged magnon-excited nonequilibrium Green function $\mathcal{G}^{\lambda}_{\mathbf{k}R(L)}(\omega,\mathbf{r})$, which takes account of the interaction with superconductor's functions $G^s_{\mathbf{k}^{\prime}L(R)}$. This interaction is calculated within the second order perturbation theory with respect to $t^{\lambda}_{R(L)\mathbf{k},\mathbf{k}^{\prime}}$. The subscripts $R$ and $L$ in $\mathcal{G}$ indicate that this function is calculated by taking into account the electron tunneling from the right, or left superconductors, respectively. Green's functions in AFM and S are represented by 2$\times2$ Keldysh matrices \cite{Rammer} given by $G_{11}=G^{\mathrm{r}},G_{22}=G^{\mathrm{a}},G_{12}=G^\mathrm{K}$ and $G_{21}=0$, where the superscripts $\mathrm{r},\mathrm{a}$ and $\mathrm{K}$ denote retarded, advanced and Keldysh functions, respectively. So, by expanding the average of Eq.(\ref{jhat})  with respect to $t^{\lambda}_{R(L)\mathbf{k},\mathbf{k}^{\prime}}$ the stationary electric current can be written in the form
\begin{eqnarray}\label{j}
&&j_{L(R)}=-e\sum_{\mathbf{k},\mathbf{k}^{\prime},\mathbf{q},\lambda}|t^{\lambda}_{L(R)\mathbf{k},\mathbf{k}^{\prime}}|^2w^L_{\mathbf{q}}w^R_{\mathbf{q}}\times \nonumber\\
&&\int \frac{d\omega}{2\pi}\mathrm{Tr}\left([\mathcal{G}^{\lambda\lambda}_{\mathbf{k}\mathbf{q}R(L)}(\omega),G^s_{\mathbf{k}^{\prime}L(R)}(\omega)]\tau_3\right)^{\mathrm{K}}\,,
\end{eqnarray}
where the trace is taken over spin and Nambu variables, while the superscript $\mathrm{K}$ denotes the Keldysh component of the Green's functions product. In the leading approximation with respect to the tunneling amplitude $G^{s\mathrm{K}}$ is given by the equilibrium function $G^{s\mathrm{K}}=(G^{s\mathrm{r}}-G^{s\mathrm{a}})\tanh(\omega/2T)$, while $\mathcal{G}^{\lambda\mathrm{K}}_{\mathbf{k}\mathbf{q}R(L)}$ is perturbed by magnons from the thermodynamic equilibrium. In the spatial representation the functions $w^{L(R)}$ describe the shapes of S-AFM interface contacts. They turn to zero outside the interfaces which have the rectangular form. For identical contacts of the width $w$, which are placed at the distance $l$ from each other, the Fourier components of  these functions are given by $w^L_{\mathbf{q}}=w^R_{\mathbf{q}}\equiv w_{\mathbf{q}}=\exp(iq_zl)q^{-1}_z\sin wq_z$.  The wavevector $q$ in Eq.(\ref{j}) is assumed to be much less than vectors $k$ and $k^{\prime}$  which are close to respective Fermi surfaces of S and AFM. In the case of a disordered interface it is reasonable to simplify Eq.(\ref{j}) by ignoring the dependence of $|t^{\lambda}_{L(R)\mathbf{k},\mathbf{k}^{\prime}}|^2$ on $\mathbf{k}$ and $\mathbf{k}^{\prime}$. Moreover, the tunneling amplitude is assumed to be independent on the sublattice index $\lambda$. Therefore, below we set $|t^{\lambda}_{L\mathbf{k},\mathbf{k}^{\prime}}|^2=|t^{\lambda}_{R\mathbf{k},\mathbf{k}^{\prime}}|^2=|t|^2$. As a consequence, the interaction with superconducting  contacts leads to the  selfenergy of $\mathcal{G}^{\lambda\lambda}_{\mathbf{k}\mathbf{q}R(L)}(\omega,\mathbf{r})$ in the form
\begin{equation}\label{Sigma}
\Sigma_{R(L)}(\omega)=|t|^2\sum_{\mathbf{k}}\tau_3G^{s}_{\mathbf{k}R(L)}(\omega)\tau_3\,.
\end{equation}
The Josephson effect stems from nondiagonal in Nambu variables terms of $\Sigma(\omega)$ which will be denoted as $\Sigma^{12}(\omega)$ and $\Sigma^{21}(\omega)$. The integration over $\mathbf{k}$ in Eq.(\ref{Sigma}) gives \cite{Kopnin}
\begin{equation}\label{Sigma2}
\Sigma^{12\mathrm{r}(\mathrm{a})}_{R}(\omega)=\frac{\pi}{i} |t|^2N_F\Delta_{R}e^{i\varphi}[(\omega\pm i0^+)^2-\Delta^2_{R}]^{-1/2}\,,
\end{equation}
where $N_F$ is the electron state density at the Fermi level of the superconductor and $\Delta_{R(L)}$ is the  magnitude of the superconducting order parameter at the right (left) contact. In the following we denote $\Delta_{R}+\Delta_L=2\Delta$. For $\Sigma^{21}_R$ and $\Sigma^{12}_L$ the sign of $\varphi$ is opposite to that in Eq.(\ref{Sigma2}).

The unperturbed impurity averaged Green function $G_{\mathbf{k}}(\omega)$ of the AFM is given by the tight binding model on a bipartite cubic lattice. N\'{e}el-ordered localized spins $S$ and conduction electrons interact  through the exchange interaction $SJ\sigma_z\lambda_z$. This model results in the electron band energies $\pm E_{\mathbf{k}}$ where $E_{\mathbf{k}}=(\epsilon^2_{\mathbf{k}}+S^2J^2)^{1/2}$ and $\epsilon_{\mathbf{k}}=-t(\cos ak_x+\cos ak_y+\cos ak_z)$,  with $t$ denoting the electron's hopping constant between nearest sites of the lattice. The disorder is created by random fluctuations $u_i$ of electron energies on lattice cites. These fluctuations are not correlated. Therefore, $\overline{u_iu_j}=\overline{u^2}\delta_{ij}$. The disorder gives rise to the finite mean elastic scattering time $1/2\tilde{\Gamma}$ of electrons, where $\tilde{\Gamma}=(1+\xi^2)\Gamma, \Gamma=\pi \overline{u^2}N_{\mu}/2$,  $\mu$ is the chemical potential  which is counted from the middle of the band, $\xi=SJ/\mu$, and $N_{\mu}$ is the state density at the Fermi level, per a unit cell. So, the retarded and advanced components of $G_{\mathbf{k}}(\omega)$ can be written, as \cite{Malshukov2}
\begin{equation}\label{G0}
G^{\mathrm{r}(\mathrm{a})}_{\mathbf{k}}(\omega)=\frac{1}{2}\sum_{\beta=\pm1}\frac{(1+\beta \hat{P}_{\mathbf{k}})}{\omega\pm i\tilde{\Gamma}+\mu\tau_3-\beta E_{\mathbf{k}}}\,,
\end{equation}
where $\hat{P}_{\mathbf{k}}=(\epsilon_{\mathbf{k}}\tau_z\lambda_x+JS\lambda_z\sigma_z)/E_{\mathbf{k}}$.

\emph{Effect of magnons}-Let us consider corrections to $G_{\mathbf{k}}(\omega)$ due to interactions with magnons. The corresponding one-particle interaction Hamiltonian has the form $H_{m}=J\mathbf{S}(\mathbf{r},t)\bm{\sigma}$, where $\mathbf{S}(\mathbf{r},t)$ is a classical field which slowly varies in space.  These variations are represented by small deviations $\mathbf{m}(\mathbf{r},t)$ of $\mathbf{S}$ from the easy axis which is parallel to the $z$ axis. So, we get $H_{m}=J(m^+\sigma^-+m^-\sigma^+)/2$, where $m^{\pm}=m_x(\mathbf{r},t)\pm im_y(\mathbf{r},t)$. Magnon's excitations in a uniaxial AFM are represented by two degenerate modes with the frequency $\Omega_{\mathbf{q}}$. In the classical regime these modes correspond to clockwise and counterclockwise precession of spins around the $z$ axis. We will consider magnetic excitations, whose lengths are larger than $l$. Therefore,
it is reasonable to set $q=0$ and $\Omega_{\mathbf{q}}|_{q=0}\equiv \Omega_0$. If the frequency of magnetic excitations produced by the nanooscillator is $\Omega<\Omega_0$, the corresponding excitations are evanescent waves. Just this regime will be considered below. So, the clockwise mode produces the first-order correction to the retarded (advanced) function  of the form
\begin{equation}\label{G1}
G^{(1)\mathrm{r}(\mathrm{a})}_{\mathbf{k},\mathbf{k}\pm\mathbf{q}}(\omega,\omega\pm\Omega)=
\frac{J}{2}G^{\mathrm{r}(\mathrm{a})}_{\mathbf{k}}(\omega)m^{\pm}\sigma^{\mp}G^{\mathrm{r}(\mathrm{a})}_{\mathbf{k}\pm \mathbf{q}}(\omega\pm \Omega)\,,
\end{equation}
where "+" and "-" signs correspond to absorption and emission of magnons, respectively. A similar expression can be written for the advanced function. In the following, since $k\simeq k_F\gg q$, where $k_F$ is the Fermi wave vector, one can set $q=0$ in Eq.(\ref{G1}). Besides Eq.(\ref{G1}), the perturbation expansion of $\mathcal{G}$  over the electron-magnon interaction contains terms which originate from electron's excitations near the Fermi surface and contribute to the Keldysh component of $\mathcal{G}$. These terms have the form
\begin{equation}\label{G1K}
G^{(1)}_{\mathbf{k}}(\omega,\omega\pm\Omega)=
\frac{\mathfrak{R}J}{2}G^{\mathrm{r}}_{\mathbf{k}}(\omega)m^{\pm}\sigma^{\mp}G^{\mathrm{a}}_{\mathbf{k}}(\omega\pm \Omega)\,,
\end{equation}
where $\mathfrak{R}$ is the renormalization factor, which is associated with diffusive transport of the electron's spin density induced by magnons (see SM). One can see from Eq.(\ref{G1}) and Eq.(\ref{G1K}) that, in contrast to diagonal with respect to spin variables nonperturbed function Eq.(\ref{G0}), the excited functions are nondiagonal. This results in triplet Cooper pairing correlations of AFM electrons.

Besides the interaction with magnons, the expression for $\mathcal{G}(\omega)$ includes selfenergy Eq.(\ref{Sigma}) which is associated with the S-AFM coupling at contacts. The interaction with contacts results in the Andreev reflection of electrons from them. In turn, the impurity averaging gives rise to the coherent scattering of electrons and Andreev-reflected holes on the same impurities. This process can be described by diffusion motion of the so correlated pairs between contacts \cite{Aslamazov}, as shown in Fig.1. Such a diffusion  takes place  at $\Gamma\gg\omega,\Omega, Dq^2$,  where $D=\overline{v_F^2}/4\Gamma$ is the diffusion constant and $\overline{v_F^2}$ denotes averaging over the Fermi surface. During  random motion electrons can flip their spins by interacting with magnons, as seen from Eqs.(\ref{G1}-\ref{G1K}). Therefore, the total spin of correlated pairs may be different from zero, which is  originally provided by the proximity to the singlet superconductor. In contrast to singlets, triplet Cooper correlations propagate over much larger distances. Formally, such diffusion processes can be expressed in terms of the diffusion propagators $\mathcal{D}_{00}^{\mathrm{r}(\mathrm{a})}(\omega)$ and $\mathcal{D}_{ij}^{\mathrm{r}(\mathrm{a})}(\omega)$ where $(i,j)\in (x,y)$ . In the absence  of the spin-orbit interaction these functions are given by \cite{Malsh}
\begin{eqnarray}\label{Dxx}
&&\mathcal{D}^{\mathrm{r}\lambda\lambda^{\prime}}_{ij}(\omega,\mathbf{q})=d^{\lambda\lambda^{\prime}}_{ij}(-2i\omega+D_{\perp}q^2)^{-1}, \nonumber \\
&& \mathcal{D}^{\mathrm{r}}_{00}(\omega,\mathbf{q})=d_0(-2i\omega+D_0q^2+4\Gamma\xi^2)^{-1}\,,
\end{eqnarray}
where $d^{11(22)}_{xx(yy)}=-\Gamma(1+\xi^2)$, $d^{12(21)}_{xx(yy)}=-\Gamma(1-\xi^2)$, $d^{11(22)}_{xy}=-d^{11(22)}_{yx}=\pm 2i\Gamma\xi$,  $d^{12(21)}_{xy}=0$, and $d_0=-\Gamma(1-\xi^2)$. The diffusion coefficients are renormalized, as $D_0= (1-\xi^2)(1+\xi^2)^{-2}D$ and $D_{\perp}= (1+\xi^2)^{-1}D$.  Expressions for advanced functions $\mathcal{D}^{\mathrm{a}}(\omega,\mathbf{q})$ are represented  by Eq.(\ref{Dxx}) with complex conjugated denominators. One can see that the spatial behavior of $\mathcal{D}^{\mathrm{r}}_{ij}(\omega,q_z)$ is determined by its pole at $q_z \simeq  (2i\omega/D)^{1/2}$, by taking into account that $q_x=0$ and in a thin film, whose thickness is much less than $l$, $q_y=0$. Since $\omega\sim\Delta_{R(L)}$, this pole corresponds to the long-range diffusion of triplet correlations, on distances comparable to a paramagnetic metal. In contrast, the diffusion of singlets, which is given by $\mathcal{D}^{\mathrm{r}}_{00}$, decreases much faster when $\xi\sim1$, at the length $|q_z|^{-1}\sim (D_0 /\Gamma)^{1/2}$. It is important that long-range functions $\mathcal{D}_{ij}(\omega,\mathbf{q})$ appear only in two-magnon processes. They are absent in the linear perturbation theory which might lead to the time-dependent current. In the latter case the diffusion occurs through propagators $\mathcal{D}_{0i}$ and $\mathcal{D}_{i0}$. In the considered model of AFM they are zero, even if the spin-orbit interaction is taken into account \cite{Malsh}.

\emph{Total Green function}-The total AFM Green functions  $\mathcal{G}^{\lambda\lambda}_{\mathbf{k}\mathbf{q}}(\omega)$ can be obtained by combining diffusion propagators from Eq.(\ref{Dxx}) with selfenergy Eq.(\ref{Sigma}) and Green's functions given by Eqs.(\ref{G0}-\ref{G1K}). It should be taken into account that, if $|t^{\lambda}_{L(R)\mathbf{k},\mathbf{k}^{\prime}}|^2$ does not depend on wavevectors and sublattice variables, the AFM Green's function enters in Eq.(\ref{j}) as the sum $\sum_{\mathbf{k}\lambda}\mathcal{G}^{\lambda\lambda}_{\mathbf{k}\mathbf{q}}$. Most simple expressions may be written for  $\mathcal{G}^{\mathrm{r}(\mathrm{a})}_{\mathbf{k}\mathbf{q}}$, because the latter involves only retarded (advanced) functions from Eq.(\ref{G1}), while $\mathcal{G}^K$ contains  products of $G^\mathrm{r}$ and $G^\mathrm{a}$ (see SM). The retarded function enters Eq.(\ref{j}) as
\begin{eqnarray}\label{Gkqr}
&&\frac{\overline{u^2}}{2}\sum_{\mathbf{k}\lambda}\mathcal{G}^{\lambda\lambda\mathrm{r}}_{\mathbf{k}\mathbf{q}L(R)}(\omega)=4\mathcal{D}^{\mathrm{r}}_{00}(\omega,\mathbf{q})
\Sigma^{\mathrm{r}}_{L(R)}(\omega)+ \sum_{\nu,\lambda\lambda^{\prime}} M^{\mathrm{r}ji}_{\lambda\lambda^{\prime}}\times\nonumber \\
&&\mathcal{D}^{\mathrm{r}\lambda^{\prime}\lambda}_{ij}(\omega+\nu\frac{\Omega}{2},\mathbf{q})
\left(\Sigma^{\mathrm{r}}_{L(R)}(\omega)+\Sigma^{\mathrm{r}}_{L(R)}(\omega+\nu\Omega)\right)\,,
\end{eqnarray}
where $\nu=\pm$ and the matrix $M$ takes account of the electron-magnon interaction. Within the diffusion approximation it does not depend on $\omega$ and $q$. By considering the dependence of the formfactor $w^{L(R)}_{q_x}$ on $l$, it becomes evident from Eq.(\ref{Dxx}) that, at $4\Gamma\xi^2 \gg D/l^2$, the integration over $q_x$ in Eq.(\ref{j}) results in a fast decreasing with $l$ of the first term in Eq.(\ref{Gkqr}). This term is associated with the singlet Cooper's correlations. At the same time, the second term which is associated with triplet correlations decreases much slower, as $\exp(-l/\kappa)$, where $\kappa\sim\min[D/\omega)^{1/2},D/\Omega)^{1/2}]$ is of the order of the coherence length $l_c\equiv\sqrt{D/\Delta}$ of superconducting correlations in nonmagnetic metals.

\emph{Evaluation of magnon's effect}-The coupling constant which characterizes the magnon's effect can be evaluated by calculating in Eq.(\ref{Gkqr}) the sum $R=\sum_{\lambda\lambda^{\prime}}M^{\mathrm{r}ji}_{\lambda\lambda^{\prime}}d^{\lambda^{\prime}\lambda}_{ij}$.  The corresponding analytical expression for $R$ at $4\xi^2\Gamma\gg \Delta$ can be obtained from SM in the form
\begin{equation}\label{M}
R=\frac{|m|^2}{8}\left(\frac{J}{\xi^2\Gamma}\right)^2(1-\xi^2)^4\,.
\end{equation}
The amplitude of the N\'{e}el order perturbation by the coupling to the nanooscillator is evaluated in SM, as $|m|=2.6\cdot 10^{-2}$, though actually its magnitude may vary in a wide range depending on material parameters and the set up. At $\xi=0.5$ and $J/\Gamma =10$ Eq.(\ref{M}) gives $R=0.04$. Since the current through a long enough junction is mostly provided  by magnon-assisted transport of triplets, it is important to gain some insight into the magnitude of such a current. It is instructive to compare it with the case when the weak link is represented by a nonmagnetic disordered metal. In this situation one may set $\xi=0$ in the lower line of  Eq.(\ref{Dxx}). As a result, dephasing term $4\Gamma\xi$ vanishes from the denominator of $\mathcal{D}_{00}$. So, the singlet pairing in Eq.(\ref{Gkqr}) may provide the long-range Josephson effect having the similar dependence on $l$, as the magnon-induced current due to the second term in Eq.(\ref{Gkqr}). However, besides this term one should take into account other contributions into the total nonequlibrium current. Its ratio to the current through the normal weak link (normalized current) can allow to evaluate the strength of the magnon effect.
\begin{figure}[tp]
\includegraphics[width=6cm]{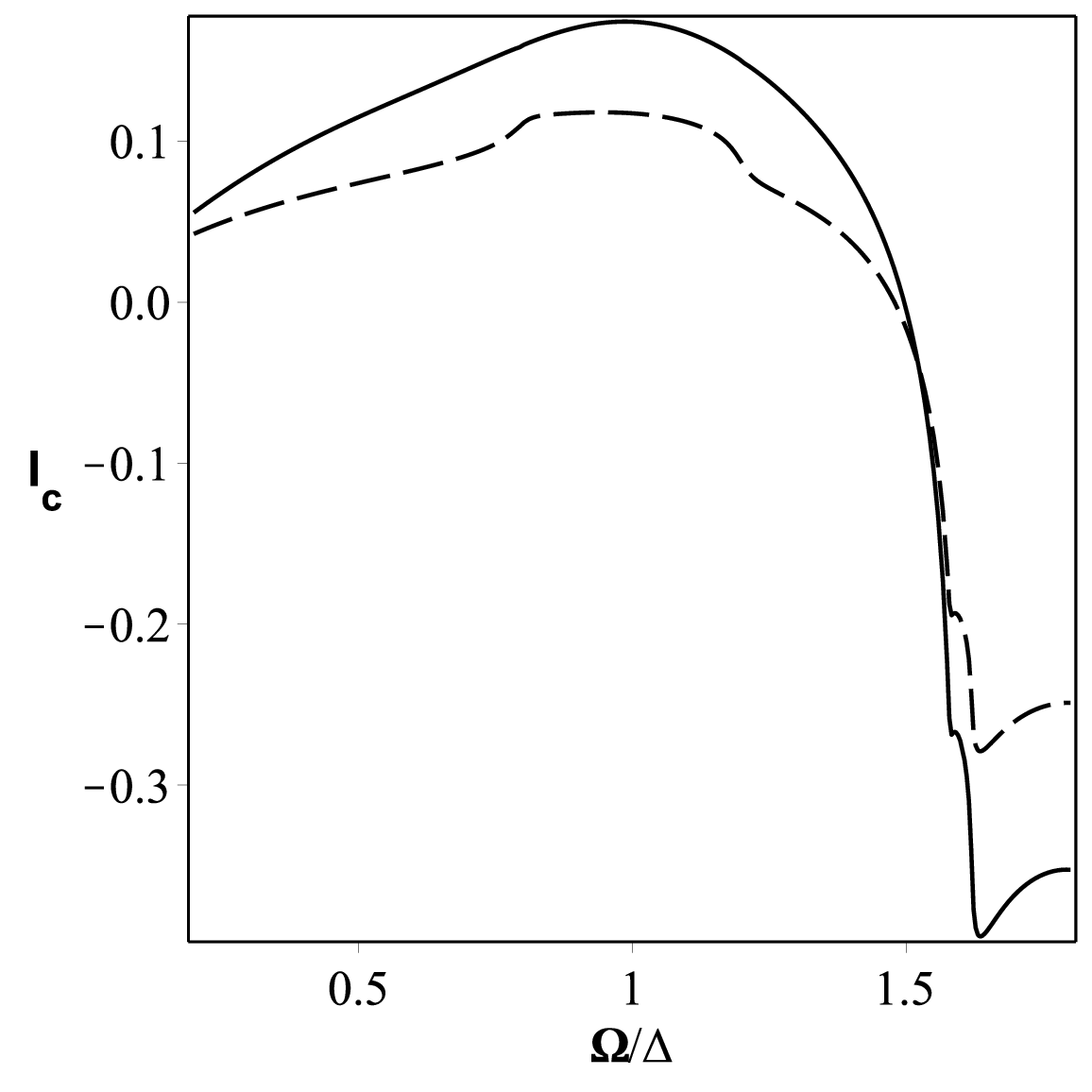}
\caption{The normalized Josephson critical current, as a function of the magnetic excitation frequency, at $k_BT=0.1\Delta$ (solid) and 0.01$\Delta$ (dashed), $l=l_c$, $\Delta_R=0.8\Delta$, $\Delta_L=1.2\Delta$} \label{fig2}
\end{figure}

\emph{Numerical results}- Fig.2 depicts the dependence of the normalized  critical current on the frequency of evanescent magnetic waves which are excited in AFM due to its contact with an antiferromagnetic nanooscillator (see the corresponding set up in Fig.S2 in SM). We set $\Omega_0=4\Delta$. Therefore, at $\Delta=$1meV $\Omega_0$ is about 1THz. Only the magnon-induced part of the current is shown at Fig.2, while the unperturbed singlet current is very small in the range of $l> l_c$. For instance, at $l=l_c$, $\xi=0.5$, and $k_BT=0.01\Delta$ the  normalized singlet current is only 6$\cdot$10$^{-3}$, that is much less than the current in Fig.2 at $\Omega=1.8\Delta$. As seen from Fig.2, the magnon effect does not considerably vary  with the temperature. At the same time, its frequency dependence is rather nontrivial. The critical current changes sign near $\Omega=1.5\Delta$. Such a 0-$\pi$ transition is also takes place at the varying junction length, as seen  in Fig.3. It is important that the frequency of spin precession generated by a nanooscillator in Ref.\cite{Khymyn} can be widely varied by the electric current in the Pt film. Hence, in this way one can also control the Josephson current and switch the JJ into the 0-$\pi$ transition regime.
\begin{figure}[tp]
\includegraphics[width=6cm]{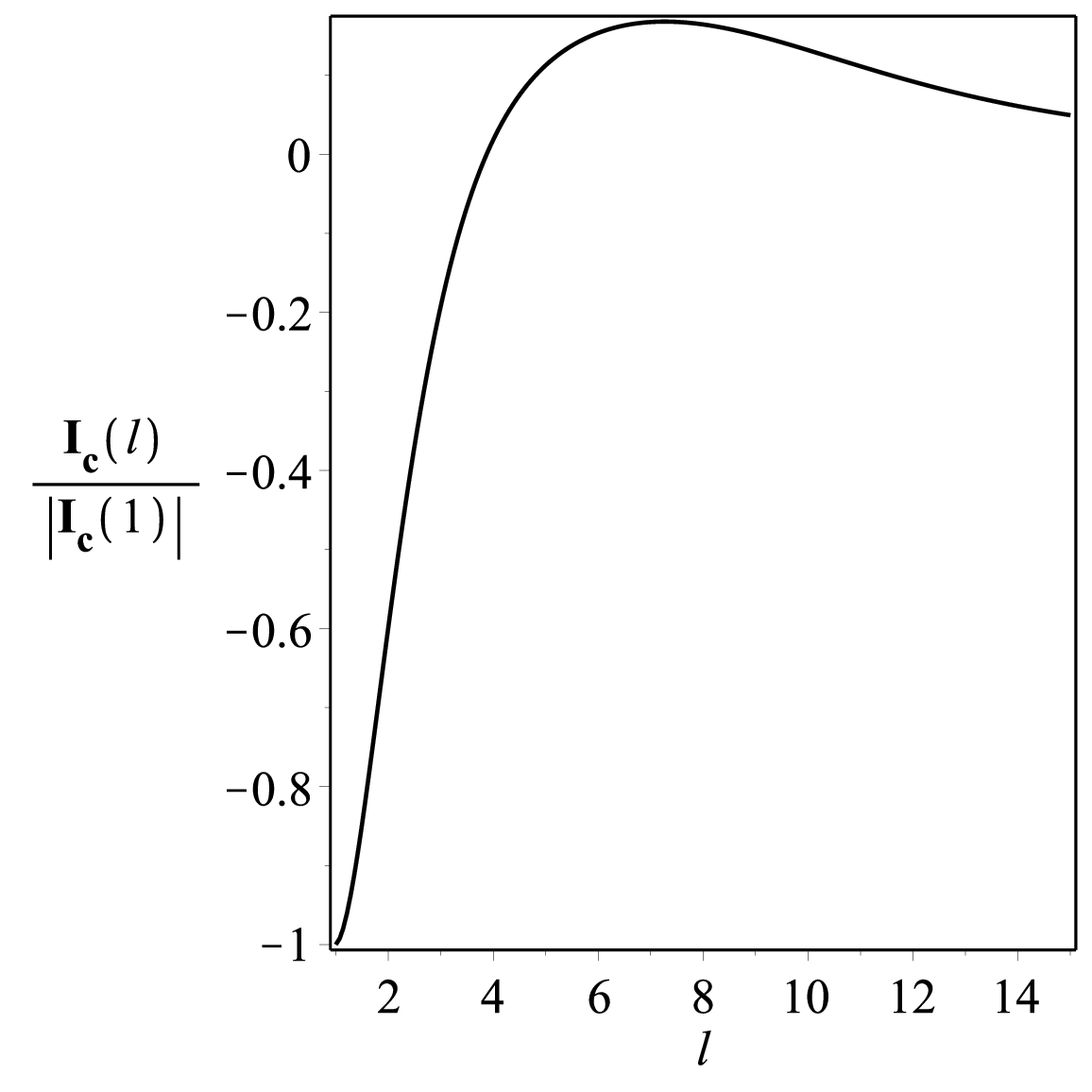}
\caption{The Josephson critical current, as a function of the junction length (in units of the coherence length $l_c$), at $\Omega$=1.8$\Delta$ and $k_BT/\Delta$=0.01} \label{fig3}
\end{figure}

\emph{Conclusion}- The nonequilibrium Green function theory was employed for the calculation of the stationary Josephson current which can be induced by a coherent magnetic excitation of the N\'{e}el order in a disordered AFM weak link. It is shown that such an excitation dramatically enhances the current through a planar junction whose length is larger than the mean free path of electrons in the AFM film. The key effect of magnons is that they can transfer their orbital moment to correlated singlet electron pairs, thereby transforming them into triplet pairs. In contrast to singlets, the latter can penetrate deep into AFM. The magnetic excitation of the N\'{e}el order can be created by a spin-Hall nanooscillator \cite{Khymyn,Cheng}, which makes a contact with the AFM film. On the other hand, magnons might be created and transported to the junction over a long distance by various means \cite{Cornelissen,Lebrun}. However, the employed here method should be modified, if these magnons have wavelengths shorter than $l$. When magnons are created by the spin-Hall nanooscillator, their polarization is controlled by its parameters. In the classical approach above only the unidirectional circular polarization was considered. At the same time, it might be elliptical or linear. In this case an extended analysis of the magnon effect is needed.


\pagebreak
\widetext
\begin{center}
\textbf{\large Supplemental Materials for "Magnon-induced transparency of a disordered antiferromagnetic Josephson junction}
\end{center}
\setcounter{equation}{0}
\setcounter{figure}{0}
\setcounter{table}{0}
\setcounter{page}{1}
\makeatletter
\renewcommand{\theequation}{S\arabic{equation}}
\renewcommand{\thefigure}{S\arabic{figure}}
\renewcommand{\bibnumfmt}[1]{[S#1]}
\section{S1. Calculation of the Josephson current}

To avoid a confusion, in this Supplemental Material the references to equations in the main text will be indicated as Eq.(...), while equations in this text are numbered as Eq.(S...), where "..." is their number.

\subsection{1. Critical current}

Let us simplify Eq.(\ref{j}) of the main text by assuming that $|t^{\lambda}_{L(R)\mathbf{k},\mathbf{k}^{\prime}}|^2=|t|^2$. So, this parameter is the same for both contacts and does not depend on wave vectors in superconducting contacts and AFM, as well as on sublattice variables of the AFM crystal. Therefore, one may integrate over $\mathbf{k}$ the Green's functions of the superconductor at the left and right contacts in Eq.(\ref{j}). As a result, Eq.(\ref{j})  can be written in the form
\begin{equation}\label{js}
j_{L(R)}=e|t|^2\sum_{\mathbf{k},\mathbf{q}\lambda}w^L_{\mathbf{q}}w^R_{\mathbf{q}}
\int \frac{d\omega}{2\pi}\mathrm{Tr}\left([\mathcal{G}^{\lambda\lambda}_{\mathbf{k}\mathbf{q}R(L)}(\omega),\Sigma_{L(R)}(\omega)]
\tau_3\right)^{\mathrm{K}}\,,
\end{equation}
where selfenergy $\Sigma_{R(L)}(\omega)$ is given by Eqs.(\ref{Sigma},\ref{Sigma2}), with $G^s_{R(L)\mathbf{k}}(\omega)$ represented by  the Keldysh matrix. The AFM Green function in Eq.(\ref{js}) should be averaged over disorder. In the Born approximation, which is valid when the electron's Fermi wave length is much smaller than the mean free path, the averaging procedure is reduced to the calculation of the so called ladder perturbation series \cite{Rammer}, as shown in Fig.(S1). Elements of the ladders are represented by two-particle loops which were calculated in Ref.\cite{Malsh} for a thermodynamically equilibrium  AFM weak link. In the considered here nonequilibrium case these two-particle functions have the form
\begin{equation}\label{Pij}
\Pi^{\mathrm{r}(\mathrm{a}),\lambda\lambda^{\prime}}_{ij,12,\mathbf{q}}(\omega+\omega^{\prime})=
\frac{u^2}{2}\sum_{\mathbf{k}}\mathrm{Tr}[G^{\mathrm{r}(\mathrm{a}),\lambda\lambda^{\prime}}_{11,\mathbf{k}+\mathbf{q}}(\omega)
\sigma^jG^{\mathrm{r}(\mathrm{a}),\lambda^{\prime}\lambda}_{22\mathbf{k}}(\omega^{\prime})\sigma^i]\,,
\end{equation}
where Green functions are given by Eq.(\ref{G0}), while $i,j\in (0,x,y)$ and subscripts "11" and "22" in Green functions denote their Nambu components, which correspond to electron and hole subspaces, respectively. The subscript "12" in $\Pi$ should be substituted for "21", if the order of these functions in their product is changed. When the electron elastic scattering rate $\Gamma\gg \omega, v_Fq$, where $v_F$ is the Fermi velocity, the summation of the ladder series results in diffusion propagators, which are given by  Eq.(\ref{Dxx}).
\begin{figure}[bp]
\includegraphics[width=6cm]{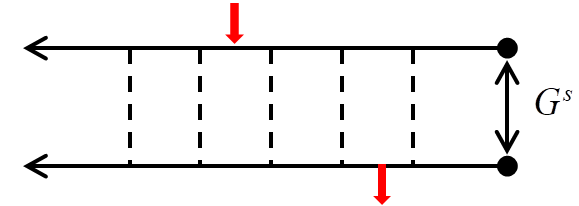}
\caption{Color online. One of the Feynman's diagrams for the disorder averaged anomalous Green function.  Dashed lines show the pair correlation function of the disorder potential. The lines connect two identical lattice sites. Red arrows indicate Green functions with corrections due to absorbed (emitted) magnons. Thick dots show the tunneling of electrons between the superconductor and AFM.} \label{fig1S}
\end{figure}
The interaction with magnons gives rise to corrections for Green functions, as  can be seen from Eqs.(\ref{G1},\ref{G1K}). These corrections appear in fermion loops, as shown in Fig.(S1). Within the second order perturbation theory there is a pair of such loops in each diagram. Some of them can be written as
\begin{equation}\label{Pi0j}
\Pi^{\mathrm{r}(\mathrm{a}),\lambda,\pm}_{0j,12,\mathbf{q}}(\omega,\Omega)=\frac{u^2}{2}
\sum_{\mathbf{k},\lambda^{\prime}}\mathrm{Tr}[\sigma^0G^{(1)\mathrm{r}(\mathrm{a}),\lambda^{\prime}\lambda}_{11,\mathbf{k}+\mathbf{q}}(\omega,\omega\pm\Omega)
\sigma^{j}G^{\mathrm{r}(\mathrm{a}),\lambda\lambda^{\prime}}_{22\mathbf{k}}(\omega)]\,
\end{equation}
and
\begin{equation}\label{Pi0jra}
\Pi^{\mathrm{raa}\lambda,\pm}_{0j,12,\mathbf{q}}(\omega,\Omega)=\frac{u^2}{2}
\sum_{\mathbf{k},\lambda^{\prime}}\mathrm{Tr}[\sigma^0G^{(1),\lambda^{\prime}\lambda}_{11,\mathbf{k}+\mathbf{q}}(\omega,\omega\pm\Omega)
\sigma^jG^{\mathrm{a},\lambda\lambda^{\prime}}_{22\mathbf{k}}(\omega)]\,,
\end{equation}
where $j\in (x,y)$, $\sigma^0$ is a unit matrix, and the functions $G^{(1)}$ in Eq.(\ref{Pi0j}) and Eq.(\ref{Pi0jra}) are given by Eq.(\ref{G1}) and Eq.(\ref{G1K}), respectively. There are also functions
$\Pi^{\mathrm{r}(\mathrm{a}),\lambda,\pm}_{j0,12,\mathbf{q}}$ and $\Pi^{\mathrm{raa}\lambda,\pm}_{j0,12,\mathbf{q}}$, where $\sigma^0$ and $\sigma^j$ have exchanged places.
Other fermion loops can be obtained by interchanging positions of $G$ and $G^{(1)}$ in Eq.(\ref{Pi0j}) and Eq.(\ref{Pi0jra}).
Such functions will be denoted as $\overline{\Pi}$. It is important that magnons always involve spin flipping of electrons. Therefore, the correlation functions given by Eq.(\ref{Pi0j}) and Eq.(\ref{Pi0jra}), as well as other correlation functions which are associated with magnons, are responsible for a singlet-triplet mixing which leads to $0j$ and $j0$ matrix elements of $\Pi$. On this reason in Eq.(\ref{Pi0j}) and Eq.(\ref{Pi0jra}) there is the summation on the sublattice index $\lambda^{\prime}$. That is because $\Pi_{00}$ in Eq.(\ref{Pij}) does not depend on the sublattice variable. Therefore the neighbors of this singlet loop in ladder series can be independently summed up on this variable.

Eqs.(\ref{Pi0j}-\ref{Pi0jra}) can be simplified by ignoring their dependence on $\omega,\Omega$ and $q$, because within the diffusion approximation the latter are much smaller than $\Gamma$. As a result, we arrive to a set of matrices which conveniently represent Eq.(\ref{Pi0j}), Eq.(\ref{Pi0jra}) and other similar correlators. For instance, $\Pi^{\mathrm{r},\lambda,\pm}_{0i,12,\mathbf{q}}(\omega,\Omega)\rightarrow\Pi^{\mathrm{rr}\lambda,\pm}_{\mathrm{r}0i}$ and $\overline{\Pi}^{\mathrm{r},\lambda,\pm}_{0i,12,\mathbf{q}}(\omega,\Omega)\rightarrow\Pi^{\mathrm{r}\lambda\pm}_{\mathrm{rr}0i}$. The difference between these two examples is that they are given by the Green functions products  $G^{(1)}_{11}G_{22}$ and $G_{11}G^{(1)}_{22}$, respectively. Other examples are $\Pi^{\mathrm{raa},\lambda\pm}_{0i,12,\mathbf{q}}(\omega,\Omega)\rightarrow\Pi^{\mathrm{ra}\lambda\pm}_{\mathrm{a}0i}$ and $\overline{\Pi}^{\mathrm{rra},\lambda\pm}_{0i,12,\mathbf{q}}(\omega,\Omega)\rightarrow\Pi^{\mathrm{r}\lambda\pm}_{\mathrm{ra}0i}$.

The function $\sum_{\mathbf{k}\lambda}\mathcal{G}^{\lambda\lambda K}_{\mathbf{k}\mathbf{q}L(R)}(\omega)$ has a more complicated structure than the retarded (advanced) function given by Eq.(10). Since we are interested in a nondiagonal in the Nambu space function, let us consider its Nambu component $\mathcal{G}^{12}$ and skip the corresponding superscript together with subscripts $L(R)$. By summing the ladder series it can be written in the form
\begin{equation}\label{Gtot}
 (u^2/2)\sum_{\mathbf{k}\lambda}\mathcal{G}^{\lambda\lambda K}_{\mathbf{k}\mathbf{q}}(\omega)=S_{\mathbf{q}0}(\omega)+\sum_{1\leqslant n \leqslant 4 }S_{\mathbf{q}n}(\omega) \,,
\end{equation}
where
\begin{equation}\label{S0}
S_{\mathbf{q}0}(\omega)=[\mathcal{D}^{\mathrm{r}}_{00}(\omega,\mathbf{q})\Sigma^{\mathrm{r}}(\omega)-
\mathcal{D}^{\mathrm{a}}_{00}(\omega,\mathbf{q})\Sigma^{\mathrm{a}}(\omega)]\tanh\frac{\omega}{2k_BT}\,
\end{equation}
is the unperturbed part of the Green's function, with $D_{00}$ given by Eq.(\ref{Dxx}). The functions $S_{\mathbf{q}n}(\omega)$ include magnetic perturbation and are given by
\begin{widetext}
\begin{eqnarray}\label{GK1}
&&S_{\mathbf{q}1}(\omega)=4\sum_{\lambda\lambda^{\prime},\nu=\pm}\left[\mathcal{D}^{\mathrm{r}}_{00}(\omega,\mathbf{q})\Pi^{\mathrm{rr}\lambda,\nu}_{\mathrm{r},0i}
\mathcal{D}^{\mathrm{r}\lambda\lambda^{\prime}}_{ij}(\omega+\nu\frac{\Omega}{2},\mathbf{q})
\Pi^{\mathrm{r}\lambda^{\prime},\overline{\nu}}_{\mathrm{rr},j0}\mathcal{D}^{\mathrm{r}}_{00}(\omega+\nu\frac{\Omega}{2},\mathbf{q})\Sigma^{\mathrm{r}}(\omega+\nu\Omega)\right.-\nonumber\\
&&\left. \mathcal{D}^{\mathrm{a}}_{00}(\omega,\mathbf{q})\Pi^{\mathrm{aa}\lambda,\nu}_{\mathrm{a},0i}
\mathcal{D}^{\mathrm{a}\lambda\lambda^{\prime}}_{ij}(\omega+\nu\frac{\Omega}{2},\mathbf{q})
\Pi^{\mathrm{a}\lambda^{\prime},\overline{\nu}}_{\mathrm{aa},j0}\mathcal{D}^{\mathrm{a}}_{00}(\omega+\nu\frac{\Omega}{2},\mathbf{q})
\Sigma^{\mathrm{a}}(\omega+\nu\Omega)\right]\tanh\frac{\omega}{2k_BT}+\nonumber\\
&&4\sum_{\lambda\lambda^{\prime},\nu=\pm}\mathfrak{R}\Pi^{\mathrm{ra}\lambda,\nu}_{\mathrm{a},0i}
\mathcal{D}^{\mathrm{a}\lambda\lambda^{\prime}}_{ij}(\omega+\nu\frac{\Omega}{2},\mathbf{q})
\Pi^{\mathrm{a}\lambda^{\prime},\overline{\nu}}_{\mathrm{aa},0j}\mathcal{D}^{\mathrm{a}}_{00}(\omega+\nu\frac{\Omega}{2},\mathbf{q})\Sigma^{\mathrm{a}}(\omega+\nu\Omega)
(\tanh\frac{\omega}{2k_BT}-\tanh\frac{\omega+\nu\Omega}{2k_BT})\,;
\end{eqnarray}

\begin{eqnarray}\label{GK2}
&&S_{\mathbf{q}2}(\omega)=4\sum_{\lambda\lambda^{\prime},\nu=\pm}\left[\mathcal{D}^{\mathrm{r}}_{00}(\omega,\mathbf{q})\Pi^{\mathrm{r}\lambda,\nu}_{\mathrm{rr},0i}
\mathcal{D}^{\mathrm{r}\lambda\lambda^{\prime}}_{ij}(\omega+\nu\frac{\Omega}{2},\mathbf{q})
\Pi^{\mathrm{rr}\lambda^{\prime},\overline{\nu}}_{\mathrm{r},j0}\mathcal{D}^{\mathrm{r}}_{00}(\omega+\nu\frac{\Omega}{2},\mathbf{q})\Sigma^{\mathrm{r}}(\omega+\nu\Omega)\right.-\nonumber\\
&&\left. \mathcal{D}^{\mathrm{a}}_{00}(\omega,\mathbf{q})\Pi^{\mathrm{a}\lambda,\nu}_{\mathrm{aa},0i}
\mathcal{D}^{\mathrm{a}\lambda\lambda^{\prime}}_{ij}(\omega+\nu\frac{\Omega}{2},\mathbf{q})
\Pi^{\mathrm{aa}\lambda^{\prime},\overline{\nu}}_{\mathrm{a},j0}\mathcal{D}^{\mathrm{a}}_{00}(\omega+\nu\frac{\Omega}{2},\mathbf{q})
\Sigma^{\mathrm{a}}(\omega+\nu\Omega)\right]\tanh\frac{\omega}{2k_BT}-\nonumber\\
&&4\sum_{\lambda\lambda^{\prime},\nu=\pm}\mathfrak{R}\Pi^{\mathrm{r}\lambda,\nu}_{\mathrm{ra},0i}
\mathcal{D}^{\mathrm{r}\lambda\lambda^{\prime}}_{ij}(\omega+\nu\frac{\Omega}{2},\mathbf{q})
\Pi^{\mathrm{r}\lambda^{\prime},\overline{\nu}}_{\mathrm{rr},j0}\mathcal{D}^{\mathrm{r}}_{00}(\omega+\nu\frac{\Omega}{2},\mathbf{q})\Sigma^{\mathrm{r}}(\omega+\nu\Omega)
(\tanh\frac{\omega}{2k_BT}-\tanh\frac{\omega+\nu\Omega}{2k_BT})\,;
\end{eqnarray}

\begin{eqnarray}\label{GK3}
&&S_{\mathbf{q}3}(\omega)=4\sum_{\lambda\lambda^{\prime},\nu=\pm}\left[\mathcal{D}^{\mathrm{r}}_{00}(\omega,\mathbf{q})\Pi^{\mathrm{rr}\lambda,\nu}_{\mathrm{r},0i}
\mathcal{D}^{\mathrm{r}\lambda\lambda^{\prime}}_{ij}(\omega+\nu\frac{\Omega}{2},\mathbf{q})
\Pi^{\mathrm{rr}\lambda^{\prime},\overline{\nu}}_{\mathrm{r},j0}\mathcal{D}^{\mathrm{r}}_{00}(\omega,\mathbf{q})\Sigma^{\mathrm{r}}(\omega)\right.-\nonumber\\
&&\left. \mathcal{D}^{\mathrm{a}}_{00}(\omega,\mathbf{q})\Pi^{\mathrm{aa}\lambda,\nu}_{\mathrm{a},0i}
\mathcal{D}^{\mathrm{a}\lambda\lambda^{\prime},\nu}_{ij}(\omega+\nu\frac{\Omega}{2},\mathbf{q})
\Pi^{\mathrm{aa}\lambda^{\prime},\overline{\nu}}_{\mathrm{a},j0}\mathcal{D}^{\mathrm{a}}_{00}(\omega,\mathbf{q})
\Sigma^{\mathrm{a}}(\omega)\right]\tanh\frac{\omega}{2k_BT}+\nonumber\\
&&4\sum_{\lambda\lambda^{\prime},\nu=\pm}\mathfrak{R}\Pi^{\mathrm{ra}\lambda,\nu}_{\mathrm{a}0i}
\mathcal{D}^{\mathrm{a}\lambda\lambda^{\prime}}_{ij}(\omega+\nu\frac{\Omega}{2},\mathbf{q})
\Pi^{\mathrm{a}\lambda^{\prime},\overline{\nu}}_{\mathrm{aa},j0}\mathcal{D}^{\mathrm{a}}_{00}(\omega,\mathbf{q})\Sigma^{\mathrm{a}}(\omega)
(\tanh\frac{\omega}{2k_BT}-\tanh\frac{\omega+\nu\Omega}{2k_BT})\,;
\end{eqnarray}

\begin{eqnarray}\label{GK4}
&&S_{\mathbf{q}4}(\omega)=4\sum_{\mathbf{k}\lambda\lambda^{\prime},\nu=\pm}\left[\mathcal{D}^{\mathrm{r}}_{00}(\omega,\mathbf{q})\Pi^{\mathrm{r}\lambda,\nu}_{\mathrm{rr},0i}
\mathcal{D}^{\mathrm{r}\lambda\lambda^{\prime}}_{ij}(\omega+\nu\frac{\Omega}{2},\mathbf{q})
\Pi^{\mathrm{r}\lambda^{\prime},\overline{\nu}}_{\mathrm{rr},j0}\mathcal{D}^{\mathrm{r}}_{00}(\omega,\mathbf{q})\Sigma^{\mathrm{r}}(\omega)\right.-\nonumber\\
&&\left. \mathcal{D}^{\mathrm{a}}_{00}(\omega,\mathbf{q})\Pi^{\mathrm{a}\lambda,\nu}_{\mathrm{aa},0i}
\mathcal{D}^{\mathrm{a}\lambda\lambda^{\prime}}_{ij}(\omega+\nu\frac{\Omega}{2},\mathbf{q})
\Pi^{\mathrm{a}\lambda^{\prime},\overline{\nu}}_{\mathrm{aa},j0}\mathcal{D}^{\mathrm{a}}_{00}(\omega,\mathbf{q})
\Sigma^{\mathrm{a}}(\omega)\right]\tanh\frac{\omega}{2k_BT}-\nonumber\\
&&4\sum_{\mathbf{k}\lambda\lambda^{\prime},\nu=\pm}\mathfrak{R}\Pi^{\mathrm{r}\lambda,\nu}_{\mathrm{ra},0i}
\mathcal{D}^{\mathrm{r}\lambda\lambda^{\prime}}_{ij}(\omega+\nu\frac{\Omega}{2},\mathbf{q})
\Pi^{\mathrm{r}\lambda^{\prime},\overline{\nu}}_{\mathrm{rr},j0}\mathcal{D}^{\mathrm{r}}_{00}(\omega,\mathbf{q})\Sigma^{\mathrm{r}}(\omega)
(\tanh\frac{\omega}{2k_BT}-\tanh\frac{\omega+\nu\Omega}{2k_BT})\,,
\end{eqnarray}
\end{widetext}
where $\overline{\nu}=-\nu$ and $\mathfrak{R}$ is the vertex renormalization constant which appears in Eq.(\ref{G1K}). Its calculation is presented in the next subsection. The matrices $\Pi$ in Eqs.(\ref{GK1}-\ref{GK4}) can be calculated from Eq.(\ref{Pi0j}),Eq.(\ref{Pi0jra}) and  Eqs.(\ref{G0}-\ref{G1K}) as
\begin{eqnarray}\label{Pi}
\Pi^{\mathrm{rr}\lambda,\pm}_{\mathrm{r},0x}&=&\Pi^{\mathrm{r}\lambda,\pm}_{\mathrm{rr},0x}=im^{\pm}\frac{J\Gamma}{4\tilde{\Gamma}^2}(1-\xi^2)(1\mp\xi\lambda);\nonumber\\
\Pi^{\mathrm{rr}\lambda,\pm}_{\mathrm{r},0y}&=&\Pi^{\mathrm{r}\lambda,\pm}_{\mathrm{rr},0y}=\pm m^{\pm}\frac{J\Gamma}{4\tilde{\Gamma}^2}(1-\xi^2)(1\mp\xi\lambda);\nonumber\\
\Pi^{\mathrm{aa}\lambda,\pm}_{\mathrm{a},0x(y)}&=&\Pi^{\mathrm{a}\lambda,\pm}_{\mathrm{aa},0x(y)}=
-\Pi^{\mathrm{ra}\lambda,\pm}_{\mathrm{a},0x(y)}=\Pi^{\mathrm{r}\lambda,\pm}_{\mathrm{ra},0x(y)}=-\Pi^{\mathrm{r}\lambda,\pm}_{\mathrm{rr},0x(y)};\nonumber\\                                                                                  \Pi^{\mathrm{rr}\lambda,\pm}_{\mathrm{r},x0}&=&\Pi^{\mathrm{rr}\lambda,\mp}_{\mathrm{r},0x}=i\Pi^{\mathrm{rr}\lambda,\pm}_{\mathrm{r},y0}\,,
\end{eqnarray}
where $\xi=SJ/\mu$. By using  these constants one can calculate the parameter $F$, which is given by
\begin{equation}\label{F}
F=\Gamma^{-1}\sum_{\lambda\lambda^{\prime}}\Pi^{\mathrm{rr}\lambda,\pm}_{\mathrm{r},0i}d^{\lambda\lambda^{\prime}}_{ij}\Pi^{\mathrm{rr}\lambda^{\prime},\mp}_{\mathrm{r},j0}\,,
\end{equation}
where $d^{\lambda\lambda^{\prime}}_{ij}$ is given by Eq.(9). So, from Eq.(\ref{Pi}) it follows that
\begin{equation}\label{F2}
F=|m|^2\frac{J^2}{2\Gamma^2}(1-\xi^2)^2\,.
\end{equation}
This parameter relates to $R$ in Eq.(\ref{M}), as $R=(F/4\xi^2)(1-\xi^2)^2$. It was obtained for the combination of $\Pi$-matrices which takes place in the upper line of Eq.(\ref{GK1}). The same expression may be obtained for other combinations of matrices $\Pi$ in  Eqs.(\ref{GK1}-\ref{GK4}). Therefore, $F$ plays the role of the electron-magnon coupling parameter. Eqs.(\ref{GK1}-\ref{GK4}) can be further simplified at the large dephasing of  singlet correlations, which is associated with the impurity scattering. In this case one can ignore  in the denominator of $\mathcal{D}_{00}(\omega,\mathbf{q})$ (see Eq.(\ref{Dxx})) its dependence on $\omega$ and $q$, because the dephasing $4\Gamma\xi^2$ dominates  at $\xi\sim 1$. Hence, in this case Eq.(\ref{Dxx}) is reduced to  $\mathcal{D}_{00}=-(1-\xi^2)/4\xi^2$. By substituting in Eqs.(\ref{GK1}-\ref{GK4}) so simplified $\mathcal{D}_{00}$ and taking into account Eq.(\ref{F}) and Eq.(\ref{F2}) we arrive at
\begin{widetext}
\begin{eqnarray}\label{S1S4}
&&\sum_{n=1}^4S_{\mathbf{q}n}(\omega)=-F\frac{1-\xi^2}{\xi^2}\sum_{\nu=\pm}\left[\tilde{\mathcal{D}}^{\mathrm{r}}_{\mathbf{q}}(\omega+\nu\frac{\Omega}{2})
(\Sigma^{\mathrm{r}}(\omega+\nu\Omega)+\Sigma^{\mathrm{r}}(\omega))-
\tilde{\mathcal{D}}^{\mathrm{a}}_{\mathbf{q}}(\omega+\nu\frac{\Omega}{2})(\Sigma^{\mathrm{a}}(\omega+\nu\Omega)+\Sigma^{\mathrm{a}}(\omega))\right]\times\nonumber\\
&&\left[\frac{1-\xi^2}{2\xi^2}\tanh\frac{\omega}{2k_BT}-\mathfrak{R}(\tanh\frac{\omega}{2k_BT}-\tanh\frac{\omega+\nu\Omega}{2k_BT})\right]\,,
\end{eqnarray}
where $\tilde{\mathcal{D}}^{\mathrm{r(a)}}_{\mathbf{q}}(\omega)$ is obtained from Eq.(\ref{Dxx}) in the form
\begin{equation}\label{Dtilde}
\tilde{\mathcal{D}}^{\mathrm{r(a)}}_{\mathbf{q}}(\omega)=\Gamma(\mp 2i\omega+D_{\perp}q^2)^{-1}\,.
\end{equation}
In a thin AFM film, whose thickness is much less than the distance between contacts, one may set $\mathbf{q}=q_x$ in $\tilde{D}^{\mathrm{r(a)}}_{\mathbf{q}}(\omega)$. From Eq.(\ref{j}) and Eq.(\ref{Sigma}),  at $w^L_{q_x}=w^R_{q_x}=w_{q_x}$, the current through the left contact can be written as
\begin{equation}\label{jL}
j_L=2e\sum_{q_x}(w_{q_x})^2\int \frac{d\omega}{2\pi}(B^{11}_{q_xL}(\omega)-B^{22}_{q_xL}(\omega))\,,
\end{equation}
where superscripts in $B$ denote the corresponding Nambu elements of the matrix given by
\begin{equation}\label{B}
B_{q_xL}(\omega)=(\Sigma^{\mathrm{r}}_{L}(\omega)+\Sigma^{\mathrm{a}}_{L}(\omega))\mathcal{G}^{K}_{q_xR}(\omega)+
(\Sigma^{\mathrm{r}}_{L}(\omega)-\Sigma^{\mathrm{a}}_{L}(\omega))
(\mathcal{G}^{\mathrm{r}}_{q_xR}(\omega)+\mathcal{G}^{\mathrm{a}}_{q_xR}(\omega))\tanh\frac{\omega}{2k_BT}\,,
\end{equation}
where $\mathcal{G}_{q_xR}(\omega)=\sum_{\mathbf{k}\lambda}\mathcal{G}^{\lambda\lambda}_{\mathbf{k}q_xR}(\omega)$. Its Keldysh component $\mathcal{G}^K_{q_zR}(\omega)$ is given by Eq.(\ref{Gtot}), while retarded and advanced functions $\mathcal{G}^{r(a)}_{q_zR}(\omega)$ can be obtained from Eq.(10).

Eqs.(\ref{jL}) and (\ref{B}) are the final results of this section. They allow to express the critical current through the amplitude of the magnetic excitation $m$ which enters into the coupling constant $(\ref{F2})$. This result does not depend on a specific mechanism which might be employed for the excitation of spin waves. It is seen from Eqs.(\ref{GK1})-(\ref{GK4}), Eq.(\ref{Sigma2}) and Eq.(\ref{B}) that $B_{q_xL}^{11}=-B^{22*}_L=iC_{q_xL}(\omega)\exp(i\varphi)$, where the function $C_{q_xL}(\omega)$ is real. Therefore, $j_L$ has the typical phase dependence $j=j_{cL} \sin\varphi$, where $j_{cL}=-4e\sum_{q_x}(w_{q_x})^2\int \frac{d\omega}{2\pi}C_{q_xL}(\omega)$ is the critical current. The symmetrized critical current $j_c$ is given by $j_c=(j_{cL}+j_{cR})/2$.
\end{widetext}

\subsection{2. Normalization of the critical current}

In order to evaluate the strength of the magnon's effect, it is convenient to normalize $j_c$ by the current through a nonmagnetic junction, which has the same geometric and disorder parameters, as the AFM junction. In the case of a nonmagnetic junction the proximity effect is not suppressed, as it takes place in AFM. Therefore, the corresponding critical current has the same long-range character, as the magnon-induced current in AFM. The corresponding normalization current $j^N$ can be obtained from Eq.(\ref{jL}) and  Eq.(\ref{B}), where in the thermodynamic equilibrium $\mathcal{G}^{NK}$ is given by the  function
\begin{equation}\label{GN}
\mathcal{G}^{NK}_{q_xR}(\omega)=[\mathcal{G}^{N\mathrm{r}}_{q_xR}(\omega)-\mathcal{G}^{N\mathrm{a}}_{q_xR}(\omega)]\tanh\frac{\omega}{2k_BT}\,.
\end{equation}
In a nonmagnetic metal
\begin{eqnarray}\label{GN2}
&&\frac{u^2}{2}\mathcal{G}^{Nr(a)}_{q_xR}(\omega)=\frac{u^2}{2}\sum_{\mathbf{k}\lambda}\mathcal{G}^{N\mathrm{r}(\mathrm{a})}_{\mathbf{k}q_xR}(\omega)=\nonumber\\
&&4\mathcal{D}^{N\mathrm{r}(\mathrm{a})}_{00}(\omega,q_x)\Sigma^{\mathrm{r}(\mathrm{a})}_{R}(\omega)\,,
\end{eqnarray}
where
\begin{equation}\label{DN}
\mathcal{D}^{N\mathrm{r}(\mathrm{a})}_{00}(\omega,q_x)=-\Gamma(\mp2i\omega+Dq^2_x)^{-1}\,.
\end{equation}
This equation can be obtained from Eq.(\ref{Dxx}) at $\xi=0$. By substituting Eq.(\ref{GN}) in Eq.(\ref{GN2}) and Eq.(\ref{B}) we obtain
\begin{eqnarray}\label{B2}
&&B^N_{q_xL}(\omega)=4[\Sigma^{\mathrm{r}}_{L}(\omega)\Sigma^{\mathrm{r}}_{R}(\omega)\mathcal{D}^{N\mathrm{r}}_{00}(\omega,q_x)-\nonumber\\
&&\Sigma^{\mathrm{a}}_{L}(\omega)\Sigma^{\mathrm{a}}_{R}(\omega)\mathcal{D}^{N\mathrm{a}}_{00}(\omega,q_x)]\tanh\frac{\omega}{2k_BT}\,.
\end{eqnarray}
The current is given by Eq.(\ref{jL}) with $B_{q_xL}$ substituted for $B^N_{q_xL}$. It can be written as $j^N=j^N_c\sin\varphi$. So calculated current corresponds to that obtained earlier for normal disordered weak links \cite{Aslamazov}. Note, that the functions $\mathcal{D}^{N\mathrm{r}(\mathrm{a})}_{00}(\omega,\mathbf{q})$ in Eq.(\ref{B2}) and $\tilde{\mathcal{D}}^{\mathrm{r(a)}}_{\mathbf{q}}(\omega)$ in Eq.(\ref{Dtilde}) have the same dependence on $q$, apart from irrelevant differences in diffusion constants.

\subsection{3. Vertex correction}

The electron-magnon interaction,  which contains a product of retarded and advanced functions in  Eqs.(\ref{GK1})-(\ref{GK4}), is renormalized due to multiple scattering of particles from disorder. Since this interaction involves spins which are perpendicular to the N\'{e}el order, it is proportional to Pauli operators $\sigma_i$ with $i\in (x,y)$.  Due to the renormalization it takes the form
\begin{equation}\label{renormal}
\mathbf{m}\cdot\bm{\sigma} \rightarrow m_i\mathfrak{R}_{ij}(\Omega)\sigma_j\,,
\end{equation}
where
\begin{equation}\label{renormal2}
\mathfrak{R}_{ij}=\delta_{ij}+\mathcal{K}_{ij}\,.
\end{equation}
In the Born approximation the correlation function  $\mathcal{K}$ can be obtained from the Bethe-Salpeter equation
\begin{equation}\label{calK}
\mathcal{K}^{\lambda\lambda^{\prime}}_{ij}(\Omega)=K^{\lambda\lambda^{\prime}}_{ij}(\Omega)+ \sum_{\mu}K^{\lambda\mu}_{il}(\Omega)\mathcal{K}^{\mu\lambda^{\prime}}_{lj}(\Omega)\,,
\end{equation}
where sublattice variables are shown explicitly. The second term of this equation generates the expansion over disorder scattering of the same sort, as shown in Fig.(S1). However, an important difference takes place. The correlation function  $K^{\lambda\mu}_{il}(\Omega)$ in Eq.(\ref{calK}) involves a pair of Green's functions which both are of particle, or hole type, while Fig.(S1) corresponds to correlated scattering of electrons and Andreev holes.  Accordingly, the correlation function $K^{\lambda\mu}_{il}(\Omega)$ is given by
\begin{equation}\label{Kij}
K^{\lambda\lambda^{\prime}}_{ij}(\Omega)=
\frac{u^2}{2}\sum_{\mathbf{k}}\mathrm{Tr}[\sigma^iG^{\mathrm{r}\lambda\lambda^{\prime}}_{\mathbf{k}}(\omega+\Omega)
\sigma^jG^{\mathrm{a}\lambda^{\prime}\lambda}_{\mathbf{k}}(\omega)]\,.
\end{equation}
Both Green's functions in this equation are "11" or "22" Nambu matrix elements. By substituting in Eq.(\ref{Kij}) Green functions from Eq.(\ref{G0}) one can see that $K^{\lambda\lambda^{\prime}}_{ij}(\Omega)$ is diagonal in the vector space at $i,j\in (x,y)$ and is independent on sublattice and Nambu variables. Therefore, the latter is also valid for $\mathcal{K}^{\lambda\lambda^{\prime}}_{ij}(\Omega)$. So, Eq.(\ref{calK})  can be easy resolved and Eq.(\ref{renormal2}) gives
\begin{equation}\label{Rij}
\mathfrak{R}^{\lambda\lambda^{\prime}}_{ij}(\Omega)=-\Gamma(1-\xi^2)\delta_{ij}(i\Omega+4\Gamma\xi^2)^{-1}\,.
\end{equation}
The second term in the denominator of $\mathfrak{R}$ describes the electron's spin density relaxation. Note, that this relaxation takes place for spins oriented perpendicular to the N\'{e}el order, in contrast to perpendicular Cooper triplets, which do not relax in the absence of the spin-orbit interaction \cite{Malsh}. When $4\Gamma\xi^2 \gg \Omega$ one can retain only the relaxation term in Eq.(\ref{Rij}). As a result, the vertex renormalization constant in Eq.(8) becomes $\mathfrak{R}=-(1-\xi^2)/4\xi^2$, which coincides with $D_{00}$ from Eq.(\ref{Dxx}) in the same range of parameters.

\section{S2. Coupling of a Josephson junction to a magnetic nanooscillator}

In the previous section we obtained the expression for the Josephson current which is induced by magnetic excitations. This effect depends on their amplitude $m$ which enters into the electron-magnon coupling constant $F$ given by Eq.(\ref{F2}). There are several ways to produce coherent magnetic excitations of the N\'{e}el order. Here we consider a set up which integrates a spin-Hall antiferromagnetic nanooscillator and the AFM layer in JJ (Fig.S2). The nanooscillator is combined of a heavy metal film which pumps the spin current into a biaxial antiferromagnetic insulator. As a result, the latter enters the autooscillation mode and generates coherent spin oscillations \cite{Khymyn,Cheng}. Then, due to spin-spin exchange interaction these oscillations are transmitted to AFM magnetic moments. Most efficiently it occurs when both antiferromagnets make contact through their uncompensated surfaces. The contact interface is assumed to be located in the $x,z$ plane. We consider oscillators which were proposed in Refs.\cite{Khymyn,Cheng}. In the high excitation regime the staggered magnetization there precesses within the $x,y$ plane. However, in Ref.\cite{Khymyn} the latter is the easy plane of the nanooscillator, while in Ref.\cite{Cheng} it is perpendicular to the easy plane and the easy axis is parallel to the $z$-axis in Fig.S2. The precession frequency is controlled by the DC electric current in the heavy metal film.  In addition, weaker higher frequency harmonics also take place. The AFM is a uniaxial antiferromagnet whose easy axis is parallel to the $z$-axis. Due to the exchange interaction on the interface the spin dynamics of the nanooscillator leads to precession of  AFM localized spins around this axis. Note, that the choice of the antiferromagnetic nanooscillator in the considered set up is important for two reasons: its higher oscillation frequency and the absence of stray magnetic fields near superconducting contacts.

It is important that  ferromagnetic and antiferromagnetic nanooscillators operate at  high dc currents  in heavy metals. Therefore, Joule heating from such devices must be removed to keep cryogenic conditions in a Josephson junction. At least for the ferromagnetic oscillator such conditions have been provided in Ref.\cite{Liu}. It is reasonable to expect that such cryogenic conditions can also be realized for antiferromagnetic oscillators.
\begin{figure}[tp]
\includegraphics[width=6cm]{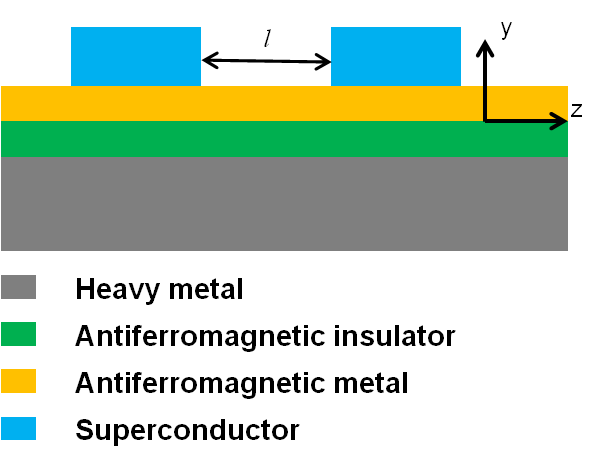}
\caption{Color online. The suggested set up for studying the effect of magnetic excitations on the Josephson current through AFM. The electric current in the Pt layer
is directed parallel to the $x$ axis. The spin-Hall current is polarized along the $z$ axis} \label{fig2S}
\end{figure}

In order to evaluate the spin precession amplitude $m$ in  Eq.(\ref{F2}), we follow the approach used in \cite{Khymyn2}. Let $\mathbf{M}_1$ and $\mathbf{M}_2$ be magnetizations of two sublattices in AFM, such that $M_1=M_2=M_s$. As usual, let us introduce the dimensionless staggered magnetization $\mathbf{l}=(\mathbf{M}_1-\mathbf{M}_2)/2M_s$ and magnetization field $\mathbf{m}=(\mathbf{M}_1+\mathbf{M}_2)/2M_s$, where $M_s$ is the saturation magnetization.  Similar fields $\mathbf{l}^{\prime}$ and $\mathbf{m}^{\prime}$ are introduced for sublattice magnetizations in the nanooscillator. These fields  are coupled on the  interface at $y=0$ via the interface exchange interaction $E_s(\mathbf{m}+\kappa \mathbf{l})\cdot(\mathbf{m}^{\prime}+\kappa^{\prime}\mathbf{l}^{\prime})$, where $\kappa$ and $\kappa^{\prime}$ are phenomenological parameters. So, when these parameters are equal to 1, both interfaces are uncompensated. At the same time, when they turn to 0 both interfaces are fully compensated.  Since for typical antiferromagnets $m\ll l$, the strongest interface exchange interaction takes place for fully uncompensated interfaces.

For small oscillations of $\mathbf{m}$ around the $z$ axis the equation of motion for $\mathbf{l}$ is given by
\begin{equation}\label{l}
\nabla^2_tl_i-c^2\nabla^2_zl_i+\Omega^2_rl_i=0\,,
\end{equation}
where $i\in (x,y)$, $c$ is the spin wave velocity and $\Omega_r$ is the frequency of the AFM resonance. The Gilbert damping has been neglected in this equation. Such an approximation is valid, as long as the width of the antiferromagnetic resonance, which is caused by this damping, is smaller than the difference between nanooscillator  and AFM resonance frequencies. The solution of this equation has the form $l_i=A_i\exp(z/\lambda)+B_i\exp(-z/\lambda)$, where $\lambda^{-1}=c^{-1}(\Omega_r^2-\Omega^2)^{1/2}$. In the considered regime of $\Omega < \Omega_r$  $\lambda$ is real and the magnetic excitation in AFM is the evanescent  wave. The boundary conditions on the free surface $(z=d)$ and on the interface with the antiferromagnetic insulator $(z=0)$ are  given by
\begin{eqnarray}\label{bc}
&c^2\nabla_z l_i|_{z=d}=&0\,, \nonumber \\
&c^2\nabla_z l_i|_{z=0}=&l^{\prime}_i\frac{\gamma E_sE_{ex}}{2M_s}\,,
\end{eqnarray}
where $E_{ex}$ is the exchange interaction  between AFM sublattices and $\gamma$ is the gyromagnetic ratio.  Only leading terms have been retained in Eq.(\ref{bc}), while the terms $\sim \nabla_tl_i/E_{ex}$ were neglected. We will take into account only the first harmonic with the frequency $\Omega$ in oscillations of $l^{\prime}_i(t)$. So, according to \cite{Khymyn} $l^{\prime}_x(t)=\cos\Omega t$, $l^{\prime}_y(t)=\sin\Omega t$ and $l^{\prime}_z(t)=0$. The feedback effect of the AFM film on the performance of the nanooscillator has been ignored, although it could be important if $\Omega \rightarrow \Omega_r$. By denoting $l^{+}=l_x+ il_y$ we obtain from Eq.(\ref{l}) and Eq.(\ref{bc})
\begin{eqnarray}\label{l2}
&&l^+(z)=\frac{\gamma E_sE_{ex}}{2 M_s\lambda(1-\exp(2d/\lambda))}\frac{1}{\Omega^2-\Omega_r^2}\times\nonumber\\
&&(e^{z/\lambda}+e^{2d/\lambda}e^{-z/\lambda})\,.
\end{eqnarray}
In antiferromagnets the magnetization field $\mathbf{m}$ can be expressed via the staggered field $\mathbf{l}$ as
\begin{equation}\label{m}
\mathbf{m}=\frac{1}{E_{ex}}(\mathbf{l}\times \nabla_t \mathbf{l}).
\end{equation}
Since in the leading approximation $l_z\simeq 1$, Eq.(\ref{m}) gives
\begin{equation}\label{mplus}
m^+(z)=-\frac{\Omega}{E_{ex}}l^+(z).
\end{equation}
This expression can be averaged over $z$, because excited spins are coupled to triplet Cooper correlations which, due to diffusion, are homogeneously distributed over thickness of a thin AFM film.
So, the averaged expression for $m^+$ is obtained from Eq.(\ref{l2}) and Eq.(\ref{mplus}) in the form
\begin{equation}\label{mfin}
m^+=\frac{\Omega E_s\gamma}{2dM_s(\Omega^2_r-\Omega^2)}.
\end{equation}
Finally, it is convenient to express $\mathbf{m}$ in terms of the spin density $\mathbf{m}M_s/\gamma=\mathbf{m}S/V_0$, where $S$ is the saturation spin of lattice sites and $V_0$ is the the unit cell volume. So, we obtain
\begin{equation}\label{deltaS}
m^+S=\frac{\Omega E_sV_0}{2d(\Omega^2_r-\Omega^2)}.
\end{equation}
To evaluate $|m|$, let us set $\Omega$=2meV, $\Omega_r$=4meV, $E_s$=20meV/nm$^2$, $d$=4nm, $S=1$, and $V_0$=(0,4nm)$^3$. At these parameters we obtain from  Eq.(\ref{deltaS}) $|m|\thickapprox$2.6$\cdot10^{-2}$.

It is important that the antiferromagnetic insulator NiO was chosen in Refs. \cite{Khymyn} and \cite{Cheng} as a part of the nanooscillator. In this case, if the hard axis of NiO is parallel to the $z$ axis in Fig(2S), its interface with AFM is magnetically compensated. Therefore, the device which was considered in \cite{Khymyn} can only weakly interact with AFM, because $l^{\prime}$ in Eq.(\ref{bc}) should be substituted for much smaller $m^{\prime}$. On the other hand, in the case considered in Ref.\cite{Cheng} the uncompensated interface magnetization, whose easy axis is parallel to $z$, might take place.
\end{document}